\documentclass[twocolumn,amsmath,amssymb]{revtex4}

\def\gsim{ \lower .75ex \hbox{$\sim$} \llap{\raise .27ex \hbox{$>$}} }

\def\lsim{ \lower .75ex \hbox{$\sim$} \llap{\raise .27ex \hbox{$<$}} }

\usepackage{graphicx}

\usepackage{dcolumn}

\usepackage{bm}

\input epsf

\begin{document}

\title{Designing Cyclic Universe Models}

\author{Justin Khoury$^{1}$, Paul J. Steinhardt$^2$ and Neil Turok$^3$}

\affiliation{
$^1$ ISCAP, Columbia University, New York, NY 10027, USA \\
$^2$Joseph Henry Laboratories, 
Princeton University, 
Princeton, NJ 08544, USA \\ 
$^3$ DAMTP, CMS, Wilberforce Road, Cambridge, CB3 0WA, UK}

\begin{abstract}

The phenomenological constraints on the scalar field potential
in cyclic models of the universe are presented. We show
that cyclic models require a comparable degree of tuning to
that needed for inflationary models. The 
constraints are reduced to a set of simple design rules including ``fast-
roll" parameters analogous to the ``slow-roll" parameters in inflation. 
 
\end{abstract}


\maketitle


In conventional big bang/inflationary cosmology \cite{guth}, 
the universe begins with the big bang and expands forever.   
The cyclic model \cite{st} is an
 alternative in
which the bang is replaced by a transition to an earlier phase of evolution. 
The history of the universe is periodic, and 
the key events that shape the large scale structure of the observable universe 
occurred a cycle ago. 
Each cycle consists of: (i) a hot big bang phase during which 
large-scale 
structures form, (ii) a phase of slow, accelerated expansion~\cite{quint}, as 
observed today, which dilutes inhomogeneities and flattens 
the universe,(iii) a phase of contraction
during which nearly scale invariant 
density perturbations are generated, and (iv) a 
big crunch/big bang transition at which matter and radiation are created
and the next cycle is triggered.
The cyclic model thus addresses the homogeneity, 
flatness and monopole problems of the 
standard hot big bang 
picture, and also provides
a nearly scale invariant spectrum of  
density fluctuations, without invoking a period of high energy 
inflation. 


The distinctive, non-inflationary mechanism for generating 
density perturbations in
cyclic models results in 
a key observational difference: whereas 
inflation predicts a nearly scale invariant spectrum 
of gravitational waves, the cyclic model does not. 
The cyclic universe model is an extension  of
 the ekpyrotic
scenario~\cite{khoury,khouryb,seiberg}, in which the
hot big bang is viewed as the result of a 
collision between two brane worlds, 
in the simplest case between two orbifold fixed planes.
The theory can be described by an effective 4d action in 
which the size of the orbifold is represented by a scalar field $\phi$ 
and the force 
between the boundary planes is represented by  
an effective potential $V(\phi)$. 
The cyclic model corresponds to regularly repeating collisions with 
a period of dark energy domination during each cycle. The
condition that cycles repeat and that the
resulting solution is an attractor requires that
$V(\phi)$ takes the general form shown in Fig.~1.

The purpose of this Letter is to summarize
the constraints 
for designing successful cyclic models. We show    
that a wide range of scalar field effective potentials 
are phenomenologically viable. 
The 
constraints on the steepness of the potential turn out to be
remarkably similar to those 
on the 
flatness of the potential in inflation. 
The constraints depend
strongly on the 
amplitude of the 
growing mode density perturbation
propagating across the bounce into the hot big
bang phase.
We employ here the recent treatment of the 
transition as a collision between branes in five dimensions~\cite{tolley},
which resolves ambiguities present in earlier 
treatments~\cite{khouryb,durrer}.
The gravity wave spectrum constraints quoted here are 
derived in~\cite{boyle}.

At all times except around the big crunch/big bang transition, 
the dynamics of the cyclic model are well described 
by the 4d effective Einstein-frame Lagrangian  
\begin{eqnarray} 
{\cal L} = \sqrt{-g}\left(\frac{{\cal R}}{2}- 
\frac{(\partial\phi)^2}{2}-V(\phi) - \beta^4(\phi)(\rho_M + \rho_R) \right)\,, 
\end{eqnarray} 
where $g$ is the determinant of the metric $g_{\mu \nu}$, ${\cal R}$ the 
corresponding Ricci scalar, and we use units where $8\pi G=1$.
The coupling
$\beta(\phi)$ depends on the details of the theory, but,
when the branes approach each other,
the warping of the extra dimension
becomes irrelevant and one finds a universal behavior
$\beta \rightarrow {\rm exp}(-\phi/\sqrt{6})$ as $\phi \rightarrow -\infty$. 
This limiting form ensures that the matter ($\rho_M$) and radiation ($\rho_R$) 
energy densities remain finite at the big crunch/bang~\cite{seiberg}. 
\begin{figure} 
\epsfxsize=2.8 in \centerline{\epsfbox{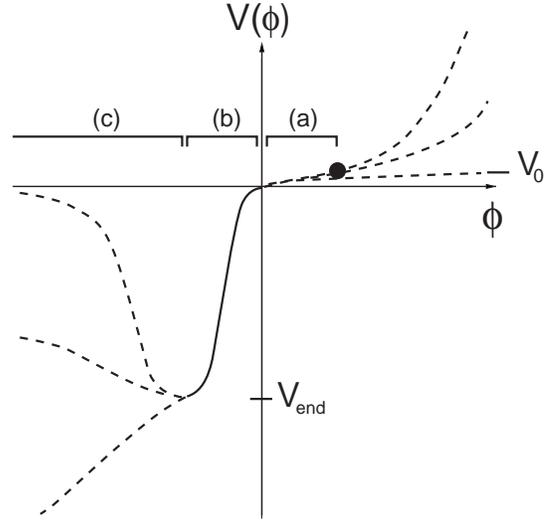}} 
\caption{Examples of cyclic potentials.} 
\label{fig:1} 
\end{figure}

The potential $V(\phi)$ in the cyclic model,
sketched in Fig.~\ref{fig:1}, can be divided into three 
regions: (a) the range where $V>0$, (b) the range where $V<0$ is nearly 
exponential and steep, (c) the range where $V$ deviates from 
this exponential steepness.
Dotted lines indicate different allowed shapes for $V$, 
illustrating the large freedom in the form of $V$.

At the present epoch, the field lies at $V=V_0$ (indicated by a dark circle) and 
its potential energy is the dark energy driving today's
observed acceleration of the 
universe (Region (a)). 
After a period of acceleration dilutes away 
inhomogeneities and flattens the  universe,
the field rolls down towards negative $V$, and 
the acceleration terminates.
Eventually, as the potential becomes 
negative, Einstein-frame expansion reverses to 
contraction (Region (b)). The large negative curvature of the
potential in this region leads to an instability amplifying 
quantum mechanical vacuum fluctuations in $\phi$ into a classical,
nearly scale 
invariant spectrum of perturbations. As Einstein-frame
contraction speeds up, $\phi$ runs off to
minus infinity in finite time 
(Region (c)) and the universe undergoes a 
big crunch/big bang transition. The brane collision 
generates 
matter and 
radiation, with the fluctuations produced in Region (b) being
imprinted as curvature perturbations in the ensuing hot big bang
phase, a process described in detail from the 5d
point of view in Ref.~\cite{tolley}.
After the collision, as the branes separate $\phi$ rebounds from
minus infinity and 
due to Hubble damping, settles somewhere in the 
range $V>0$. 
After the standard radiation- and matter-dominated epochs,
the positive potential energy once
again triggers cosmic acceleration. 
The cycle then repeats itself.

As indicated in Fig.~\ref{fig:1}, there is tremendous flexibility in  
the shape of $V(\phi)$ in Regions (a) and (c). 
Most of the quantitative constraints apply to Region (b), where 
density perturbations are generated. 
Even here,
once four primary constraints (labeled $(i)$-$(iv)$ below) 
are satisfied, the remaining requirements which we
discuss are generally satisfied as well.

{\bf Region (b)}: {\it Spectral index of density perturbations}. 
Scale invariant fluctuations can be generated during a contracting phase if the 
energy density in $\phi$ dominates, and if its equation of state 
parameter is much greater than unity and nearly constant, as shown
by Gratton {\it et al.}~\cite{scaleinv}. 
A nearly scale invariant spectrum is obtained if $V<0$ and 
$\epsilon \equiv (V/V_{,\phi})^2 \ll 1$ and $|\eta|  
\equiv |1-(V V_{,\phi \phi}/V_{,\phi}^2)| \ll 1$, 
where $V_{,\phi}=dV/d\phi$. 
Qualitatively, these require, respectively, that $V$ be negative, 
very steep and 
nearly exponential in form. 
These are almost the exact opposite of the slow-roll conditions of inflation 
which require $V$ to be positive and flat. 
In particular, $\epsilon$ and $\eta$ should be thought of as ``fast-roll'' 
parameters, in analogy with the inflationary slow-roll parameters: 
$\bar{\epsilon}\equiv V_{,\phi}^2/2V^2$ and $\bar{\eta}\equiv V_{,\phi\phi}/V$.

The fast-roll conditions imply that the cosmological evolution is described 
by~\cite{scaleinv}
\begin{equation} 
a\sim (-\tau)^{2\epsilon}\,,\qquad H \approx - \sqrt{-2 \epsilon V}\,,
\qquad\phi' \approx \sqrt{4\epsilon}/\tau\,, 
\label{back} 
\end{equation} 
where $\tau<0$ is conformal time and prime denotes $\tau$ derivative.
Since $\epsilon\ll 1$, the scale factor varies very slowly during this phase. 
Moreover, it follows from Eqs.~(\ref{cancel}) that the kinetic and 
potential energy of $\phi$ nearly cancel out:
\begin{equation}
\frac{\phi'^2}{2a^2} \approx -V\,.
\label{cancel}
\end{equation}

 
As in inflationary cosmology, the spectral index can be neatly expressed in 
terms of $\epsilon$ and $\eta$. 
Using Eq.~(\ref{back}),
the equation for the Fourier mode $k$ 
of the gauge-invariant variable $u$, related to 
the Newtonian potential $\Phi$ by $u = a\Phi/\phi'$,  reduces to
\cite{scaleinv}
\begin{equation} 
u_k''+ \left[k^2 - \frac{2(\epsilon + \eta)}{\tau^2}\right]u_k = 0\,, 
\label{eq:bessel2b} 
\end{equation} 
from which one can read off the spectral index
\begin{equation} 
n_s - 1 = -4(\epsilon + \eta)\,. 
\label{eq:nsek} 
\end{equation} 
The analogous equation for inflation leads to the familiar result 
$n_s - 1 =-6\bar{\epsilon} + 2\bar{\eta}$.

In deriving Eq.~(\ref{eq:nsek}), it is assumed that $\epsilon$ and $\eta$ are 
nearly constants for observable modes. 
For a spectral range of $N\sim 60$ e-folds, it is straightforward to show that 
constancy  
of $\epsilon$ requires $|\eta| < 1/4N$. 
Combining this with the observational constraint $|n_s-1|\;\lsim\; 0.1$, we find 
\begin{equation} 
(i)\qquad \epsilon \;\lsim\; 1/40\,, \qquad |\eta| \;\lsim\; 1/240\,. 
\label{eq:condsek2} 
\end{equation} 
The same analysis for inflation results in  
nearly identical  constraints, 
$\bar{\epsilon}\;\lsim \;1/60$ and $\bar{\eta}\;\lsim \;1/120$.

{\it Amplitude of density perturbations}. Assuming that fluctuations in $\phi$ 
start in their Minkowski vacuum, when
$k^2\tau^2\gg 2(\epsilon+\eta)$,
the solution to Eq.~(\ref{eq:bessel2b}) in the long-wavelength limit is
\begin{equation}
k^{3/2}\Phi_k \approx 2^{-3/2}\left(\frac{\phi'}{a}\right)\left(\frac{-
k\tau}{2}\right)^{-2(\epsilon +\eta)}\,.
\label{eq:Phiek2}
\end{equation}
This expression holds provided $\epsilon$ and $\eta$ are small, which we
assume is true until $\phi$ approaches the point $\phi_{end}$ where
$V=-V_{end}$ in Fig.~\ref{fig:1}. The last factor is of order unity and
only weakly $k$-dependent for small $\epsilon$ and $\eta$, hence we ignore
it for the rest of this section.
Using Eq.~(\ref{cancel}) to rewrite $\phi'/a$ in
terms of $V$, we find $k^{3/2}\Phi_k \approx
V_{end}^{\frac{1}{2}}/2$ at $\phi \approx \phi_{end}$.
                                                                                
As $\phi$ moves past $\phi_{end}$, its evolution becomes
kinetic-dominated and the potential becomes irrelevant (Region c)).
In this kinetic-dominated phase, the fractional energy density
perturbation on comoving slices for small $k$, 
$\epsilon_m \equiv 2k^{2} \Phi_k/ 3H^2
a^2 $, is equal to a constant, $\epsilon_0$,
 all the way to the
singularity $\tau=0$.  Hence, it is sufficient to 
evaluate $\epsilon_m$ at the onset
of the kinetic-dominated phase, when $\phi\approx\phi_{end}$, to obtain
\begin{equation}
\epsilon_0 = \frac{k^2V_{end}^{1/2}}{3H^2_{end}a^2_{end}}.
\label{epsend}
\end{equation}
%
Eq. (2) implies
 $H^2_{end} = 2\epsilon V_{end}$. Moreover, adopting the normalization
of the scale factor in Ref.~\cite{tolley} (see Sec. IIIC), we have
$a^2_{end} \approx v_{coll}^{2/3}L^{-2/3}\epsilon^{-1/3}V_{end}^{-1/3}$,
where $v_{coll}$ is the relative speed (assumed nonrelativistic) of the
two colliding branes and $L$ is the curvature scale associated with the
warping of the extra dimension (in heterotic M-theory
models~\cite{wittenlukas}, for example, typical values lie in the range $L
= 10^{4-6}$). Substituting in Eq.~(\ref{epsend}), we find:
\begin{equation}
k^{3\over 2} \epsilon_0 \approx {k^2 L^{2\over 3} \over 3 V_{end}^{1\over
6} \epsilon^{2\over 3} v_{coll}^{2\over 3}}.
\label{eq:Phiek2b}
\end{equation}

From Ref.~\cite{tolley},
the long wavelength
curvature perturbation on comoving slices
following the boundary brane collision is then given,
for non-relativistic $v_{coll}$,  by
\begin{eqnarray}
 k^{3/2} \zeta \sim \frac{3  v_{coll}^4 \epsilon_0}{32 L^2 k^{1/2}} \sim
 \frac{1}{32} \, \frac{v_{coll}^{10/3}   }{\epsilon^{2/3}  L^{4/3}
 V_{end}^{1/6}}.
 \label{eq:zetaexp3}
 \end{eqnarray}
 A second radiation-dependent contribution is also
 discussed in Ref.~\cite{tolley}, but this is always
 sub-dominant if the
 radiation density on the branes
 is small compared to the brane kinetic energies at collision
 (see
 equation
 (\ref{eqrad}) below).

Measurements of the cosmic microwave background (CMB)  
constrain $k^{3/2} \zeta$ 
to be $10^{-5}$, which implies 
\begin{eqnarray} 
V_{end}^{1/4} \approx  10^{5} v_{coll}^5 \epsilon^{-1}  L^{-2}.  
\label{cond3} 
\end{eqnarray}

{\it Bounds on the radiation density.} 
Although it is not necessarily required, we will consider the 
case where 
the radiation energy 
density produced at the brane collision 
is smaller than the magnitude of the brane 
kinetic energies ($\sim v_{coll}^2 L^{-2}$).
From the
formulae of Section IIIC of Ref.~\cite{tolley},
we infer the Hubble constant $H_r$ at radiation-kinetic
equal density,
$H_r \sim T_r^2$, obtaining 
\begin{equation} 
T_r \, \lsim \, \sqrt{v_{coll}/L}.  
\label{eqrad}
\end{equation} 
This constraint ensures that the radiation-dependent contribution 
to $\zeta$ mentioned following equation (\ref{eq:zetaexp3}) above is 
small compared to the radiation-independent 
contribution considered here.
 
The temperature at the beginning of the radiation-dominated epoch 
should exceed  
$\sim 1$~MeV to recover the successful predictions of the hot big bang
nucleosynthesis.
This implies 
\begin{equation} 
(ii) \, \, T_r \,  \approx \,  H_r^{1/2} \, > \, 10^{-21}. 
\label{eq:2} 
\end{equation} 
 
{\it Cycling constraint.} 
Once the radiation-dominated epoch begins, the scalar field kinetic energy is 
rapidly damped and the field comes to halt. 
In order to cycle, the radiation-dominated epoch 
should not begin 
until $\phi$ has time to go from $- \infty$, past $V=-V_{end}$, 
right across the potential well and back up the 
potential to the plateau where $V=V_0$.    
During this period, the scalar field kinetic energy  
density is dominant and $\phi- \phi_{end} \sim \sqrt{2/3} \ln(t/t_{end})$, 
where 
$t_{end}$ is the time it takes $\phi$ to reach $V=-V_{end}$ after the brane 
collision.  Assuming an exponentially steep potential 
as in Ref.~\cite{st} with $V \approx V_0 (1-{\rm exp}(-c 
\phi))$ (corresponding to $\epsilon = c^{-2}$ in the potential well),
the time required to climb from $V\approx -V_{end}$ to $V_0$ and begin the  
radiation-dominated epoch is $t_r > t_{end} (V_{end}/V_0)^{\sqrt{3/2}/c}$. 
Equality between scalar field kinetic energy and radiation occurs at 
$t_r/t_{end} \approx V_{end}^{1/2}/H_r $. Consequently, we have the constraint 
\begin{equation} 
(iii) \, \, T_r \, \lsim \, V_{end}^{1/4} (V_{0}/V_{end})^{\sqrt{3 \epsilon/8}}. 
\label{eq:3} 
\end{equation} 
 
{\it Gravitational wave bound.} 
The calculation of the spectrum of gravity waves is very similar to that for 
scalar perturbations presented above. 
A systematic analysis is given in Ref.~\cite{boyle}.  
The most  stringent requirement is that the gravitational wave contribution to 
the total energy density of the universe be less than 10\% of the radiation 
density at nucleosynthesis; this results in the constraint 
\begin{equation} 
(iv) \, \, T_r \, \gsim \, 0.1 \,  \epsilon V_{end}. 
\label{eq:4} 
\end{equation}

{\it Spectral range of the fluctuations}.  
The perturbations produced in the ekpyrotic phase as $\phi$ rolls from  
$V \approx 0$ to $\approx -V_{end}$ is  
$k_{max}/k_{min} \approx \sqrt{V_{end}/V_0}$.
To span 60 e-folds, for example, we require 
\begin{equation} 
V_0^{1/4}< e^{-30}V_{end}^{1/4} \approx 10^{-13}V_{end}^{1/4}\,. 
\end{equation} 
This does not explain why the dark energy is so small, but 
it does imply that 
an exponential hierarchy of scales is necessary in order for
the perturbations to be nearly scale invariant over 
a broad range of scales.
 
As illustrated in Figure 2, 
a very broad 
range of parameters satisfies all of the 
conditions in Region (b). 

{\bf Region (a)}: In the range $V>0$, 
there is a lot of freedom for different designs. 
First, the value of $V$ today ($V_0$), must equal $10^{-120}$ in order 
to account for the observed dark energy. 
This fine-tuning should not be viewed as a disadvantage relative to inflation 
since the dark energy must be explained in that scenario as well. 
Second, the potential must join onto the steep negative part by either a 
plateau, as in Fig. 1, or an energy barrier as 
will be discussed elsewhere. In the former case, the 
potential must be sufficiently flat for the current acceleration to be  
sustained for several tens of e-foldings, enough to smooth out the universe 
and reduce the perturbations to a level insignificant 
compared to the new fluctuations
produced in the subsequent contracting phase. 
If the number of e-foldings is greater than  
60 (about 3 trillion years),  the current particle 
density ($10^{80}$ particles per 
Hubble volume) is reduced to less than one per Hubble volume.
This is more than enough to ensure
the universe is made homogeneous, isotropic and flat. 
The condition is relatively easy to satisfy, as illustrated 
in most quintessence 
models~\cite{quint}. 
The form of $V$ for $\phi$ greater than the current value is
not important for the phenomenology of the cyclic model,
although it may be in determining whether the cyclic solution
is a global attractor.
Third, $\phi$ must not violate precision tests of general relativity. 
In the case of the plateau where the field is nearly massless, this can 
be avoided by having the couplings to matter obey
$d \ln \beta/d \phi < 10^{-3}$ for $\phi$ near the current 
value~\cite{st}.   
In the case of an energy barrier, the field is currently trapped at a local 
minimum and its mass can easily 
be made sufficiently large to avoid equivalence principle
violations on observable scales.

{\bf Region (c)}: The only constraint in this region is  
that the kinetic energy in $\phi$ must dominate the 
potential energy near the big crunch/big bang transition 
($\phi \rightarrow -
\infty$), a necessary condition for a bounce according  
to the prescription in Ref.~\cite{seiberg}. 
This will be the case if $a^6 V \rightarrow 0$ as $\phi \rightarrow -\infty$. 
For example, potentials in which  
$V(\phi)$  approaches  a constant or monotonically decreases (more slowly 
than $1/a^6$) as 
$\phi \rightarrow - \infty$ satisfy this condition.  
An example is  $V(\phi) \propto -e^{-c\phi}$ with $c< \sqrt{6}$. 
The M theory setup motivates one to consider\cite{st}
potentials where  $V \rightarrow 0$ as $\phi\rightarrow -\infty$  
since M theory reduces to a weakly coupled string theory in this limit.
However there does not seem to be any such requirement
imposed from the phenomenology.

\begin{figure} 
\epsfxsize=2.8 in \centerline{\epsfbox{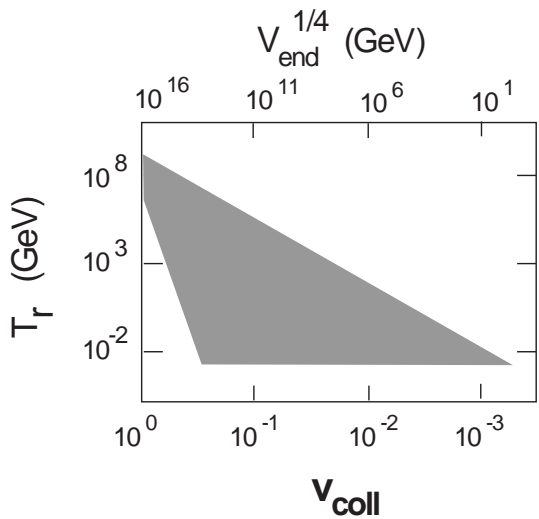}} 
\caption{The shaded region shows the range 
$V_{end}^{1/4}$, the reheat temperature 
$T_r$, and the relative speed of the colliding
branes $v_{coll}$ over which
the cyclic model 
satisfies all known cosmological constraints, with 
fixed $\epsilon=10^{-2}$ and $L=3 \times 10^4$ (in units where 
$8\pi G=1$).
Note that the reheat temperature 
is well below the grand unified monopole mass in all cases.} 
\label{fig:2} 
\end{figure} 
 
The tightest constraints on the model emerge from
Region (b), 
where density perturbations are created.
Similarly, in 
inflationary cosmology,   only the region of the potential involved
in fluctuation generation is 
tightly constrained.

For practical purposes, only the four constraints 
labeled $(i)$-$(iv)$ impose any  
significant restrictions on cyclic model building.   
Constraint $(i)$ (Eq.~\ref{eq:condsek2})  
imposes a constraint on the steepness of the potential  
 expressed through  fast-roll parameters $\epsilon$ 
and $\eta$, that are analogous to inflationary slow-roll parameters.
Constraints $(ii)$-$(iv)$ (Eqs.~\ref{eq:2}, \ref{eq:3}, and \ref{eq:4})  
limit the remaining physical parameters.  
 
As illustrated 
in Fig.~2,  a very 
wide range of physically plausible parameters is allowed,
broadly similar to that for inflationary
models. Hence in the absence of deeper theoretical 
constraints on the form of potentials, and parameters such
as the brane speeds at collision, it is premature to
claim a significant advantage for either model in terms of tuning.

We thank L. Boyle, U. Seljak and A. Tolley for helpful discussions. 
This  work is supported in part by the CU Academic Quality Fund  
and the Ohrstrom Foundation (JK), by PPARC in the UK (NT) and by  
US Department of Energy grant DE-FG02-91ER40671 (PJS).

\nopagebreak

\end{document}